\documentclass[manuscript]{aastex}

\received{}
\revised{}
\accepted{}

\shorttitle{LMC star clusters}
\shortauthors{Piatti, A.E.}

\begin{document}

\title{CCD SDSS $gr$ photometry of poorly studied star clusters in the Large
Magellanic Cloud}

\author{Andr\'es E. Piatti}
\affil{Observatorio Astron\'omico, Universidad Nacional de 
C\'ordoba, Laprida 854, 5000, C\'ordoba, Argentina\\
Consejo Nacional de Investigaciones Cient\'{\i}ficas y T\'ecnicas, Av. Rivadavia 1917, C1033AAJ,
Buenos Aires, Argentina
}
\email{e-mail: andres@oac.uncor.edu}

\begin{abstract}

We present for the first time CCD SDSS $gr$ photometry, obtained at the Gemini 
South telescope with the GMOS attached,
of stars in the field of the poorly studied star clusters NGC\,1768, HS\,85, SL\,676,
NGC\,2107, NGC\,2190, and SL\,866, which are distributed in the main body of
the Large Magellanic Cloud.
We applied a subtraction procedure to statistically clean the cluster CMDs from field star 
contamination. In order to disentangle cluster features from those belonging to their 
surrounding fields,
we applied a subtraction procedure which makes use of variable cells to reproduce
the field star Color-Magnitude Diagrams (CMDs) as closely as
possible. We then traced their stellar density radial profiles from star counts
 performed over the cleaned field stars dataset and derived their radii.
Using the cleaned cluster CMDs, we estimated ages and metallicities from matching theoretical 
isochrones computed for the SDSS system. The studied star clusters have ages from 0.1 up to 
2.0 Gyr and are of slightly metal-poor metal content ([Fe/H] $\approx$ -0.4 dex).

\end{abstract}


\section{Introduction}

The Large Magellanic Cloud (LMC) harbors more than two thousand catalogued ordinary 
star clusters (Bica et al. 2008). Although they are prime indicators of the chemical evolution 
and the star formation history of the galaxy, only a very small percentage have been 
well studied (Chiosi et al. 2006, Glatt et al. 2010). In this sense, detailed investigations 
of even a handful of clusters 
represents a significant improvement of our knowledge of the chemical enrichment
history of this galaxy. 

We have been intensively involved in a long-term project aimed at obtaining ages and metallicities
of LMC clusters, as well as addressing other important related issues. For instance,
we have    
discovered a new giant branch clump structure (Piatti et al. 1999),
studied the infamous cluster age-gap (Piatti et al. 2002),
searched for age and metallicity gradients (Piatti et al. 2009),
derived ages
and metallicities for some 81 LMC clusters (Piatti et al. 2011a, Piatti 2011), 
and investigated in detail the LMC field and cluster Age-Metallicity Relationships (Piatti 
\& Geisler 2013), among others.
We continue here our previous work on LMC clusters by
presenting results for six mostly unstudied clusters (NGC\,1768, HS\,85, SL\,676,
NGC\,2107, NGC\,2190, and SL\,866) with
the aim of adding them to our growing sample of well-studied LMC clusters that will allow
us to assemble a much more comprehensive database with which to
study the formation and evolution of LMC clusters and their parent galaxy.

The paper is organized as follows.
The next section describes the collected observations and the data reduction. 
Sect. 3 deals with the observed Color-Magnitude Diagrams (CMDs) and the procedure
of disentangling cluster from field star features. We focus also on the estimation of
the cluster structural parameters. The cluster fundamental parameters are derived 
in Sect. 4, while the analysis and discussion of the results are
presented in Sect. 5. Our main findings are summarized in Sec. 6.

\section{Data handling}

Based on data obtained from the Gemini Science Archive, 
we collected CCD SDSS $gr$ (Fukugita et al., 1996) images centerd on 6 LMC clusters
(GS-2010B-Q-74, PI: Pessev) along with observations of standard
fields and calibration frames (zero,
sky-flat, dome-flat). The data were obtained at the Gemini South telescope with the 
Gemini Multi-Object Spectrograph (GMOS) attached (scale = 0.146 arcsec/pixel). 
The log of the observations is presented in Table 1, where the main astrometric, photometric 
and observational information is summarized.
Nine Gemini Observatory standard fields 
were observed along the 5 cluster observing nights as baseline observations, for which 
2 exposures of 5 s per filter and 
airmass in the range $\sim$ 1.0-2.0 were obtained.

The data reduction followed the procedures documented in the Gemini Observatory 
webpage\footnote{http://www.gemini.edu}
and utilized the {\sc gemini/gmos} package in IRAF\footnote{IRAF is distributed by the National 
Optical Astronomy Observatories, which is operated by the Association of 
Universities for Research in Astronomy, Inc., under contract with the National 
Science Foundation.}. We performed overscan, trimming, bias subtraction, flattened all data images, etc., 
once the
calibration frames (zeros and flats) were properly combined.
The final field of view of the images resulted to be $\sim$ 5.6' $\times$ 5.6'.

Around 30-50 independent magnitude measures of standard stars
were derived per filter using
the {\sc apphot} task within IRAF, in order to secure the transformation
from the instrumental to the SDSS $gr$ standard system.  Standard stars were
distributed over an area similar to that of the GMOS array, so that we measured magnitudes of standard stars in 
each of the three chips.  The relationships between
instrumental and standard magnitudes were obtained by fitting
the following equations:

\begin{equation}
g = g_1 + g_{std} + g_2\times X_g + g_3\times (g-r)_{std}
\end{equation}

\begin{equation}
r = r_1 + r_{std} + r_2\times X_r + r_3\times (g-r)_{std}
\end{equation}

\noindent where $g_i$, and $r_i$ (i=1,2,3) are the fitted coefficients, and
$X$ represents the effective airmass. 
We solved the transformation equations for the three chips with the {\sc fitparams}
task in IRAF, simultaneously;
the rms errors from
the transformation to the standard system being 0.015 mag for $g$ and 0.023 for $r$, respectively, 
indicating an excellent match to the standard system.

The stellar photometry was performed using the star-finding and point-spread-function (PSF) fitting 
routines in the {\sc daophot/allstar} suite of programs (Stetson et al., 1990). 
For each frame, a quadratically varying 
PSF was derived by fitting $\sim$ 60 stars, once the neighbors were eliminated using a preliminary PSF
derived from the brightest, least contaminated 20-30 stars. Both groups of PSF 
stars were interactively selected. We then used the {\sc allstar} program to apply the resulting PSF to the 
identified stellar objects and to create a subtracted image which was used to find and measure magnitudes of 
additional fainter stars. This procedure was repeated three times for each frame. Finally, 
we computed aperture corrections from the comparison of PSF and aperture magnitudes by using the 
neighbor-subtracted PSF star sample. After deriving the photometry for all detected objects in
each filter, a cut was made on the basis of the parameters
returned by {\sc daophot}. Only objects with $\chi$ $<$2, photometric error less than 2$\sigma$ above 
the mean error at a given magnitude, and $|$SHARP$|$ $<$ 0.5 were kept in each filter
(typically discarding about 10\% of the objects), and then the
remaining objects in the $g$ and $r$ lists were matched with a
tolerance of 1 pixel and raw photometry obtained.

We combined all the independent instrumental magnitudes using the stand-alone {\sc daomatch} and 
{\sc daomaster} programs, kindly provided by Peter Stetson. As a result, we produced one dataset per cluster 
containing
the $x$ and $y$ coordinates for each star, and two ($g$,$g-r$) pairs. 
The gathered photometric information was standardized using 
equations (1) to (2). We finally  averaged 
standard magnitudes and colors of stars measured twice.
The resulting standardized photometric tables
list successively a running number per star, the $x$ and $y$ coordinates, the averaged $g$ 
magnitudes,  the observational errors $\sigma(g)$, the $g-r$ colors, the observational errors $\sigma(g-r)$, 
and the number of observations. We adopted 
the photometric errors provided by {\sc allstar}\footnote{Program kindly provided by P.B. Stetson} for stars 
with only one measure. Tables 2 to 7 provide this information for NGC\,1768, HS\,85, SL\,676, NGC\,2107,
NGC\,2190, and SL\,866, respectively. Only a portion 
of Table 2 is shown here for guidance regarding its form and content. The whole content of Tables 2-7 is provided as Supplementary Tables.

\section{Analysis of the Color-Magnitude diagrams}

In order to obtain extracted CMDs where the fiducial features of the clusters can be clearly seen,
we: (i) cleaned the cluster CMDs from the field star contamination by using
field stars placed beyond the cluster regions; (ii) determined the cluster geometrical centers and; 
(iii) traced the cluster radial profiles in order to determine the cluster extents.

As for cleaning the cluster CMDs from the field star contamination, we used the method developed
by Piatti \& Bica (2012), which is designed to statistically reproduce the respective field star CMD and then to
subtract it from the observed cluster CMD. The method is based on the fact that some parts of the field star
CMD are more populated than others so that, by counting the number of stars within boxes of a fixed size
becomes in a less profitable task. In general, bigger boxes are required to
satisfactory reproduce CMD regions with a small number of field stars, while smaller 
boxes are necessary in populous CMD regions. For instance, relatively
bright field red giants with small photometric errors can be subtracted only if large enough 
boxes are used and therefore, a cluster CMD without such a spurious red giant features can be built.
Piatti \& Bica proposed to use variable boxes in the field star CMDs. Magnitude and color box sizes are 
allowed to vary separately, and fixed in such a way that 
they result bigger in CMD regions with a small number of stars, and vice versa.
The boxes are placed and designed by taking into
 account the stellar density in the field star CMD, while the field stars are
eliminated by looking for one star -the closest one in terms of magnitude and color- in the 
cluster CMD for each  star identified in the field CMD. For our purposes, the field star CMDs were built
using stars located typically beyond 700 pixels from the cluster centers. The bottom-right panel of
Figs. 1-6 shows the resulting boxes in the field CMD.

The coordinates of the cluster centers and their estimated uncertainties were determined by fitting Gaussian 
distributions to the star counts in the $x$ and $y$ directions for each cluster. These projected
stellar densities were built using intervals of 40 pixel wide, although we checked that using spatial bins 
from 20 to 60 pixels does not result in significant changes in the derived centers.
We made use of the
{\sc ngaussfit} routine in the {\sc stsdas/iraf} package, which was executed from entering initial guesses
for the single Gaussian's parameters, namely: a fixed constant -in our case equals to zero- which represents 
 the corresponding background levels (i.e. stellar field densities assumed to be uniform), the linear terms to 
zero, the centers of the Gaussians, their amplitudes and their full width at half-maximum (FWHM).
 We iterated the fitting procedure on average once, after eliminating 
a couple of discrepant points. Cluster centers were finally determined with a typical standard deviation of 
$\pm$ 10 pixels ($\sim$ 1.5") in all cases.

The cluster radial profiles were then obtained 
by first counting the number of stars in adjacent boxes of 20 $\times$ 20 pixels covering the whole field 
of each cluster. 
Thus, at any distance $r$ from the cluster center,
we computed the mean stellar density using the equation:

\begin{equation}
(n_{r+10} - n_{r-10})/(m_{r+10} - m_{r-10}),
\end{equation}

\noindent where $n_j$ and $m_j$ represent the sum of the number of stars counted in boxes closer than $j$
to the cluster centre and the number of boxes found inside $j$, respectively. 
Note that Eq. (3) provides with the mean stellar density at a distance $j$ even though 
complete circles cannot be traced at that distance. This is an important consideration since having a stellar 
density profile which extends far away from the cluster center, allows us to estimate the background level more
precisely. 
Such profiles were in turn useful to derive the cluster radii, defined as the distance from the cluster center 
where the stellar density profile intersects the background level,  as well as to measure the FWHM of the cluster
density profiles, which play a significant role - from a stellar content point of view - in the construction 
of the cluster CMDs. 
When choosing the size of the rings we preferred 20 pixels which allows us to statistically sample the 
stellar spatial distribution as well as to avoid spurious effects mainly caused by the presence of localized groups, 
rows or columns of stars. Nevertheless, we traced the cluster radial profiles using rings with different sizes around
20 pixels wide in order to estimate the uncertainties in the resulting radial profiles. Typically, the
uncertainties vary from the center outwards with a S/N ratio between 8 and 33; the average being 14.
The resulting density profiles are shown in the upper-right panel of Figs. 1-6. 
We fitted a King (1962) model to these stellar density profiles using the expression:

\begin{equation}
N/N_o = ({\frac{1}{\sqrt{1+(r/r_c)^2}} - \frac{1}{\sqrt{1 + (r_t/r_c)^2}}})^2 + bkg
\end{equation}

\noindent where $N_o$ is the central stellar density, and $r_c$ and $r_t$ are the core and tidal radii, respectively.
$bkg$ represents the background level. $r_c$ and $r_t$ were estimated with a typical precision of 10 and 100
pixels, respectively, and their resulting mean values are listed in Table 8.

We then constructed three CMDs covering different circular extractions as shown in Figs. 1-6 (upper-left, 
bottom-left, and bottom-right panels). The upper-left panel corresponds to the observed cluster CMD, as
built from stars distributed within a circle of radius equals to the cluster radius. The bottom-left
panel depicts the resulting cleaned cluster CMD, once the decontamination of field stars was performed;
while the bottom-right panel shows a reference field star CMD built from stars distributed within an
equal cluster area. The observational errorbars are drawn on the right hand of each panel. As can be seen, 
the observed cluster CMDs exhibit as the most obvious traits 
Main Sequences (MSs) which vary in extent and in number of stars, besides the presence of Red Clump (RC) and 
Red Giant Branch (RGB) stars. In some cases, populous Sub-Giant Branches are also visible.
Note that all these features are also seen in the field star CMDs -although at a different stellar
density level-, which reflect the LMC composite stellar populations. The comparison of
the observed cluster and reference field star CMDs clearly becomes in a robust evidence that
field star decontamination is needed in order to disentangle the fiducial cluster features.

Despite the fact that some residual of the field star decontamination is unavoidable, the
cleaned cluster CMDs reveal that we are dealing with clusters spread in a relatively wide age range.
NGC\,1768, HS\,85, SL\,676, and NGC\,2107 appear to be relatively or moderately young star clusters,
whereas NGC\,2190 and SL\,866 seems to be of intermediate-age. In addition, SL\,676, NGC\,2190, and
possible NGC\,2107 show RCs with an elongated or secondary structure which resemble that of clusters
with evidence of age spread (e.g., Milone et al. 2009, Keller et al. 2012, Piatti 2013).

\section{Cluster fundamental parameters}

Based on the cleaned cluster CMDs we followed the common procedure of matching theoretical isochrones
in order the find the ones which best reproduce the fiducial cluster features. We chose the
evolutionary models developed by Marigo et al. (2008) for three different metallicities Z = 0.004, 0.008, and
0.020 ([FeH] = -0.7, -0.4, and +0.0, respectively) to evaluate the metallicity effect in the cluster 
fundamental parameters. The selected values cover the metallicty range for most of the 
LMC clusters younger than $\sim$ 4 Gyr (Piatti \& Geisler 2013). Note that cluster metallicity 
plays an important role when fitting theoretical isochrones. The 
distinction is mainly evident for the evolved RC and RGB phases. ZAMSs are often less affected by
metallicity effects and can even exhibit imperceptible variations for a specific metallicity range 
within the expected photometric errors. 

Before matching the cluster CMDs with theoretical isochrones, we need to adopt the cluster interstellar 
extinctions and distance moduli. As for the cluster distance moduli, considering the line-of-sight 
depth of the galaxy to be approximately 6 kpc (Crowl et al. 2001), and bearing in mind that any cluster 
of the sample could be placed in front of or behind the main body of the LMC, we concluded that the difference in 
the cluster apparent distance moduli could be as large as $\Delta(V-M_V)$ $\sim$ 0.15 mag, if a value of 
50 kpc is adopted for the mean LMC distance. Since $\Delta(V-M_V)$ resulted smaller than the uncertainties
when adjusting the isochrones, the simple assumption of adopting a unique value for 
the distance modulus for all the clusters should not dominate the error budget in our final results.
For this reason, we adopted for all the clusters the value 
of the LMC distance modulus $(m-M)_o$ = 18.50 $\pm$ 0.10 recently reported by Glatt et al. (2010).

The estimation of cluster reddening values was made by interpolating the extinction maps of Burstein \& 
Heiles (1982, hereafter BH). BH maps were obtained from H\,I (21 cm) emission data for
the southern sky. They furnish us with foreground $E(B-V)$ color excesses which depend on the Galactic coordinates.
We also derived the values of $E(B-V)$ provided by Haschke et al. (2011, hereafter HGD) based on photometry of RR 
Lyrae ab stars
obtained by the third phase of the Optical Gravitational Lensing Experiment (OGLE III). 
Although two cluster fields resulted
to be outside their extinction maps (NGC\,2190 and SL\,866), we found a fairly good agreement for the remaining
four star clusters of $\Delta(E(B-V)_{BH-HG}$ = (-0.026 $\pm$ 0.022) mag. We also compared the $E(B-H)_{BH}$ values
with those coming from the Schlegel et al. (1998, hereafter SFD) full-sky 100-$\mu$m dust emission maps.
However, their values deviate for star clusters located in the LMC bar or arms (Harris \& Zaritsky 2009) due to
saturation of H\,I emission. This is the case of: NGC\,1768, located in the North-West end of the Bar; HS\,85,
located in the North-West Arm and; SL\,676 and NGC\,2107 located in the South-East end of the Bar, respectively.
For NGC\, 2190 and SL\,866, which are placed in the South-Eastern and North-Eastern outer disk, respectively, 
the agreement between BH and SFD reddenings resulted satisfactory 
($\Delta(E(B-V)_{BH-SFD})$ = (-0.020 $\pm$ 0.005) mag). Table 8 lists the adopted $E(B-V)_{BH}$ color excesses.
We adopted $R$ = $A_V$/$E(B-V)$ = 3.1 to convert color excess to extinction, and
 used the equations $A_g$/$A_V$ = 1.199 and $A_r$/$A_V$ = 0.858 (Fan 1999) to evaluate the total extinctions 
in $A_g$ and $A_r$. Finally, we used $E(g - r)$/$A_V$ = 0.341 for the selective extinction in the SDSS system.

We then selected a set of isochrones, and superimposed them to the cluster CMDs, once they were 
properly shifted by the corresponding $E(g-r)$ color excesses and by the LMC distance 
modulus. In the matching procedure we used seven different isochrones for each metallicity 
level, ranging from slightly younger than the derived cluster age to slightly older. Finally, 
we adopted as the cluster age the one corresponding to the isochrone which best reproduced the cluster 
main features in the CMD, bearing in mind the observational errorbars and the errors in 
$E(g-r)$ and $(m-M)_o$ as well. The presence of RCs and/or RGBs in some cluster CMDs made the fitting procedure 
easier. We noted, however, that the theoretically computed bluest stage during the He-burning core phase 
is redder than the observed RC in the CMDs of some clusters, a behaviour already 
detected in other studies of Galactic and Magellanic Cloud clusters (e.g., Piatti et al. 2011b and
references therein). Notice that we do not provide with metallicity errors, since we
only used  three prearranged values in the isochrone matching. However, for the sake of the subsequent analysis,
the metallicity values adopted are in excellent agreement with those for LMC clusters of similar
ages (Piatti \& Geisler 2013). In Fig. 7 we plotted, for each cluster CMD, the isochrone of the adopted cluster age 
and two additional isochrones bracketing the derived age.  The ages of the 
bracketing isochrones were estimated by taking into account the observed dispersion in the cluster CMDs. 
The ages of the adopted isochrones and their corresponding metallicities for the cluster sample
are listed in Table 8.

\section{Analysis and discussion}

As far as we are aware from searching the literature, only NGC\,1768 has a previous age estimate.
Glatt et al. (2010) obtained an age of log($t$) = 7.8 $\pm$ 0.4 in fairly good agreement with our
present value, although their uncertainty is noticeably larger. Glatt et al. have used data from 
the Magellanic Cloud Photometric Surveys (Zaritsky et al. 2002) to build the cluster CMD. 
Although they mention that field contamination
is a severe effect in the extracted cluster CMDs and therefore influences the age estimates, no
decontamination from field CMDs were carried out. Their large age errors could reflect  
the composite stellar populations of the LMC Bar field towards which the cluster is projected. 

SL\,676 and NGC\,2017 resulted to be a cluster pair relatively close in age, with an age 
difference of (350 $\pm$ 210) Myr. These objects present an angular 
separation in the sky of 4.1', which is equivalent to 59.6 pc. However, since the upper 
separation limit for binary LMC star clusters is $\sim$ 20 pc (Bathia et al. 1991, Dieball et al. 2002)
we concluded that they do not constitute a physical system.

Finally, NGC\,2190 and SL\,866 resulted to be intermediate-age star clusters. According to their
positions in the galaxy, the resulting ages are in good agreement with those of star clusters
placed at a similar deprojected distance from the LMC center, whereas the present metallicties
result slightly more metal-rich for those galactocentric distances (Piatti et al. 2009). Comparing the
cluster ages and metallicities with those of their respective surrounding star fields
(Piatti \& Geisler 2013), we found that the latter
are older ($< t >$ $\sim$ 5 Gyr) and more metal-poor ([Fe/H] $\sim$ -1.0 dex). The remarkable different 
ages and metallicities of the star clusters
and the dominant field stellar populations could be explained if
we assume that the clusters were born in other parts of the galaxy
and, because of their orbital motions, they are observed at the current locations. Notice that the
ages of NGC\,2190 and SS\,866 are encompassed within the well-known star cluster bursting formation epoch
(Piatti 2011), so that they could have been formed in regions where the cluster burst took place.

\section{Summary}

In this study we present for the first time CCD SDSS $gr$ photometry
of stars in the field of poorly studied LMC star clusters, namely: NGC\,1768, HS\,85, SL\,676,
NGC\,2107, NGC\,2190, and SL\,866. The star clusters are spread throughout the Bar, Arms, and
Outer Disk of the galaxy.
The data were obtained at the Gemini South telescope with the GMOS attached. We are confident
that the photometric data yield
accurate morphology and position of the main cluster features in the CMDs.
We applied a subtraction procedure to statistically clean the cluster CMDs from field star contamination 
in order to disentangle cluster features from those belonging to their surrounding fields.
The technique makes use of variable cells in order to reproduce the field CMD as closely as
possible. We trace their stellar density radial profiles from star counts performed
over the cleaned field star datasets.
From the density profiles, we adopted cluster radii defined as the distance from the cluster center 
where the stellar density profile intersects the background level, and derived the radii at the FWHM of the
radial profile.  We then built CMDs with cluster features clearly identified.
Using the cleaned cluster CMDs, we estimated ages and metallicities from matching theoretical 
isochrones computed for the SDSS system. When adjusting a subset of isochrones we took into account the
LMC distance modulus and the individual star cluster color excesses. The studied star clusters turned out to
cover a relatively wide age range, from relatively young up to intermediate-age clusters. 
We found that SL\,676 and NGC\,2107 are not binary clusters but aligned along the same line-of-sight, while
NGC\,2109 and SL\,866 are intermediate-age and slightly metal-poor clusters located in the Outer Disk where
the dominant stellar populations are older and more metal-poor. The remarkably different ages and metallicities
could be explained if we consider the star cluster orbital motions.

\acknowledgements
This work was partially supported by the Argentinian institutions CONICET and
Agencia Nacional de Promoci\'on Cient\'{\i}fica y Tecnol\'ogica (ANPCyT).

\newpage

\begin{deluxetable}{lcccccc}
\tablecaption{Observation log of selected LMC clusters.}
\tablehead{\colhead{Star Cluster}  & \colhead{$\alpha_{\rm 2000}$}
 & \colhead{$\delta_{\rm 2000}$} & \colhead{filter} & \colhead{exposures}  
          &  \colhead{airmass} & \colhead{seeing} \\
                                   & \colhead{(h m s)}            
 & \colhead{($^o$ $'$ $"$)}      &                  
 & \colhead{(times $\times$ sec)} &                   & \colhead{"}}

\startdata

NGC\,1768 &  04 57 02 & -68 14 56  & $g$ & 2$\times$30 & 1.283 & 1.2 \\
          &           &            & $r$ & 2$\times$15 & 1.281 & 1.1 \\
HS\,85    &  05 00 51 & -67 48 14  & $g$ & 2$\times$30 & 1.308 & 0.9 \\
          &           &            & $r$ & 2$\times$15 & 1.306 & 0.9 \\
SL\,676   &  05 43 09 & -70 34 16  & $g$ & 2$\times$30 & 1.321 & 0.7 \\ 
          &           &            & $r$ & 2$\times$15 & 1.322 & 0.6 \\
NGC\,2107 &  05 43 13 & -70 38 23  & $g$ & 2$\times$30 & 1.321 & 0.7 \\
          &           &            & $r$ & 2$\times$15 & 1.322 & 0.6 \\
NGC\,2190 &  06 01 02 & -74 43 33  & $g$ & 2$\times$30 & 1.509 & 1.2 \\
          &           &            & $r$ & 2$\times$15 & 1.503 & 1.1 \\
SL\,866   &  06 14 32 & -65 58 57  & $g$ & 2$\times$30 & 1.407 & 1.1 \\
          &           &            & $r$ & 2$\times$15 & 1.402 & 1.0 \\

\enddata

\end{deluxetable}

\begin{deluxetable}{lccccccc}
\tablecaption{CCD $gr$ data of stars in the field of NGC\,1768.}
\tablehead{\colhead{Star} & \colhead{$x$}  & \colhead{$y$} & \colhead{$g$} & 
\colhead{$\sigma$$(g)$}  & \colhead{$g-r$} & \colhead{$\sigma$$(g-r)$} & \colhead{n} \\
\colhead{}     & \colhead{(pixel)} & \colhead{(pixel)} & \colhead{(mag)} & 
\colhead{(mag)} & \colhead{(mag)} & \colhead{(mag)}  & }

\startdata

-    &   -     &   -     &  -    &  -    &   -   &   -     \\
     11 &1493.480 &1187.271 &  16.663 &   0.002 &  -0.197  &  0.004&   2\\
     12 &1685.743 &1275.260 &  16.605 &   0.002 &  -0.226  &  0.004&   2\\
     13 &1874.385 & 611.421 &  16.661 &   0.002 &   0.608  &  0.003&   2\\
-    &   -     &   -     &  -    &  -    &   -   &   -     \\

\enddata

\end{deluxetable}

\setcounter{table}{7}

\begin{deluxetable}{lccccccc}
\tablecaption{Fundamental parameters for selected LMC star clusters. }
\tablehead{\colhead{Star Clusters} & \colhead{\it{l}}  & \colhead{b} & \colhead{r$_c$} & 
\colhead{r$_t$}  & \colhead{E(B-V)} & \colhead{log($t$)} & \colhead{Z} \\
\colhead{}     & \colhead{(deg)} & \colhead{(deg)} & \colhead{(pixel)} & 
\colhead{(pixel)} & \colhead{(mag)} &  & }

\startdata

NGC\,1768    & 279.360 & -35.549 & 90  & 800  & 0.05 & 8.00 $\pm$ 0.20 & 0.008 \\
HS\,85       & 278.716 & -35.325 & 160 & 1000 & 0.06 & 8.65 $\pm$ 0.10 & 0.008 \\
SL\,676      & 281.126 & -31.125 & 90  & 500  & 0.07 & 8.80 $\pm$ 0.10 & 0.008 \\
NGC\,2107    & 281.205 & -31.114 & 120 & 1300 & 0.07 & 8.45 $\pm$ 0.10 & 0.008 \\
NGC\,2190    & 285.768 & -29.408 & 200 & 2000 & 0.12 & 9.10 $\pm$ 0.10 & 0.008 \\
SL\,866      & 275.776 & -28.332 & 210 & 2000 & 0.05 & 9.30 $\pm$ 0.10 & 0.008 \\
\enddata

\end{deluxetable}

\begin{figure}[htb]
\includegraphics[angle=0,width=17cm]{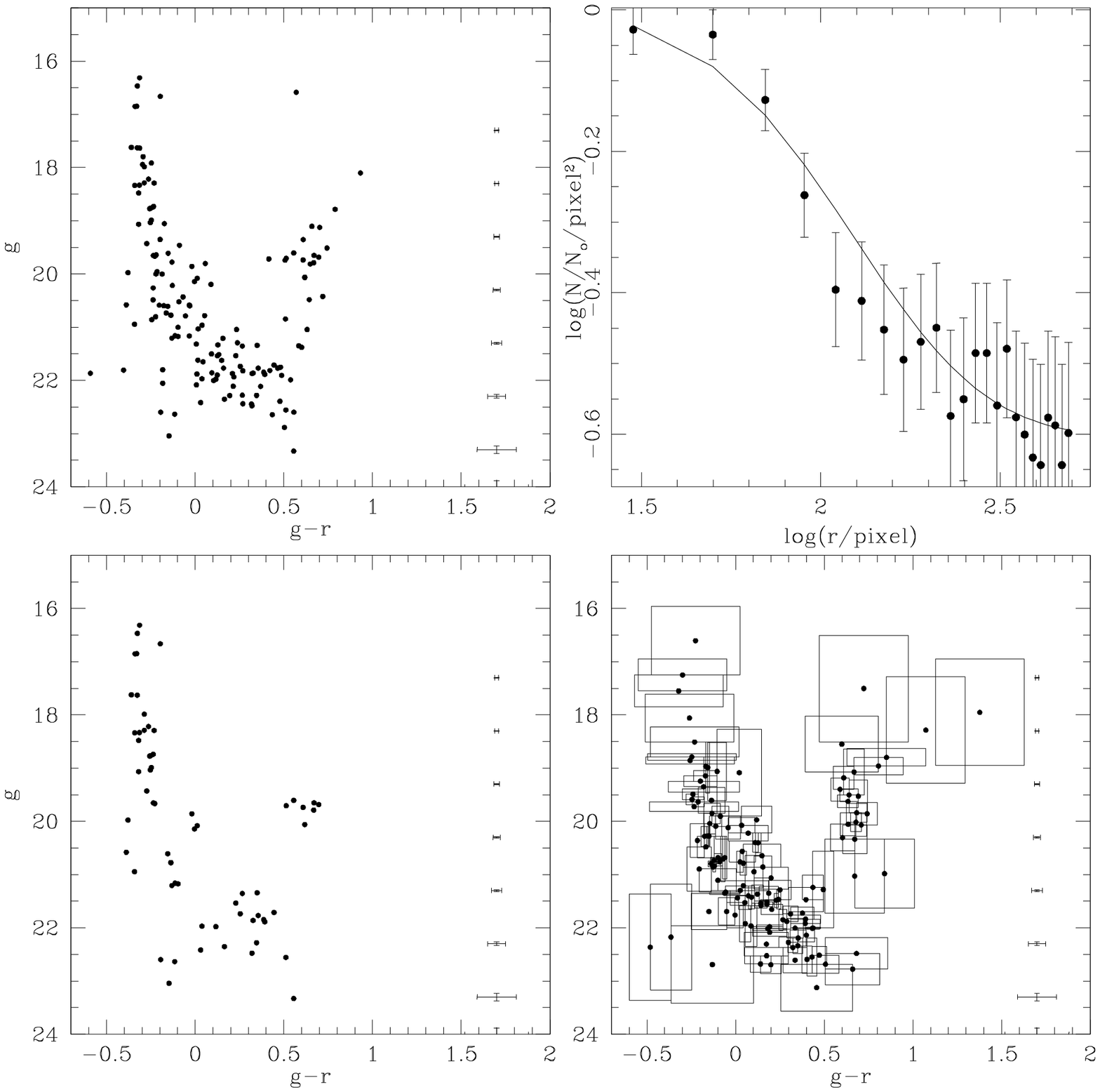}
\caption{Three
extracted CMDs for stars measured in the field of NGC\,1768 distributed within the cluster radius (upper-left panel), 
the cluster surrounding field for an equal cluster area (bottom-right panel), and the cluster cleaned from field 
contamination (bottom-left panel). The cluster radial profile is also depicted (upper-right panel).}
\end{figure}

\begin{figure}[htb]
\includegraphics[angle=0,width=17cm]{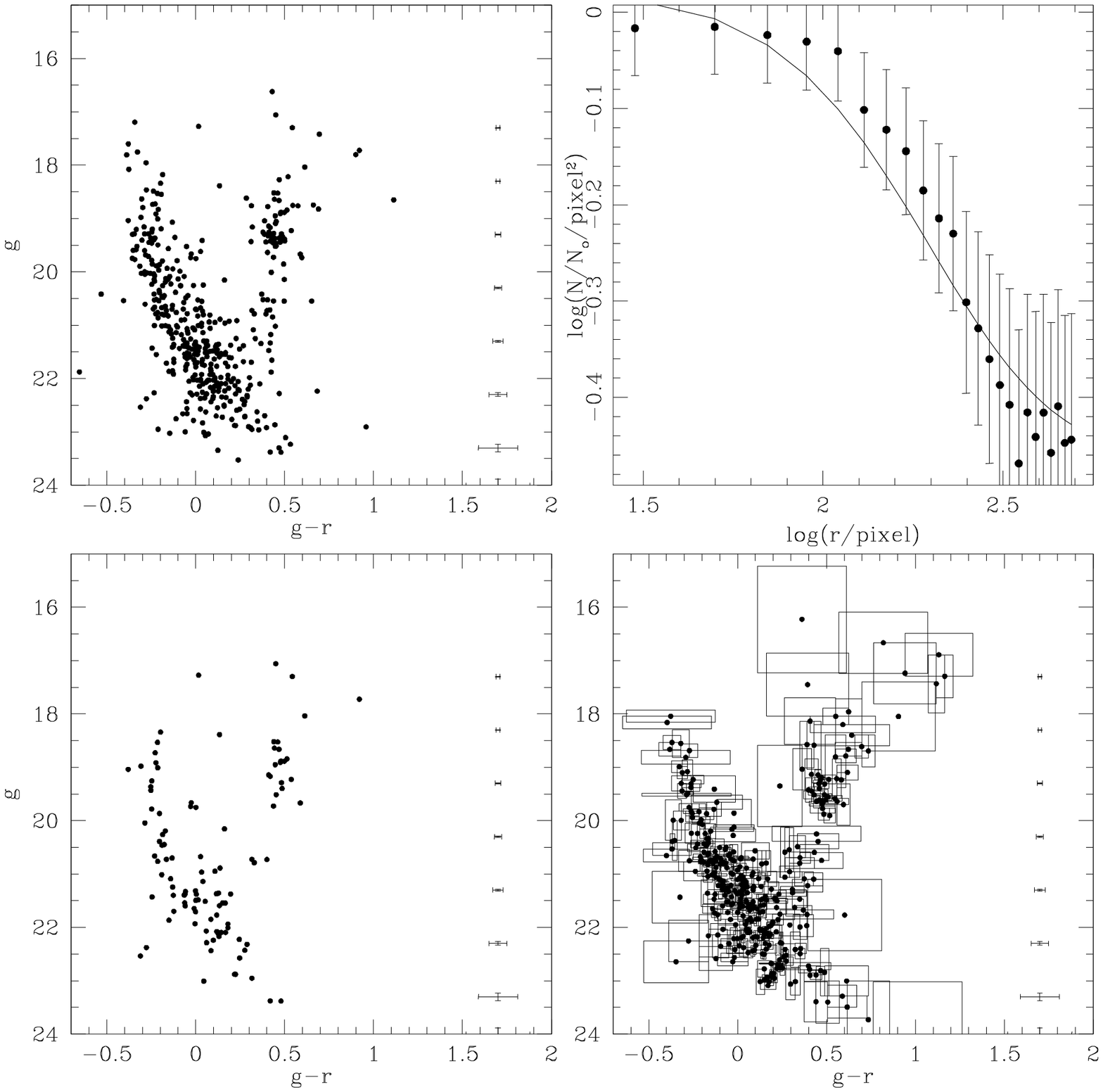}
\caption{Three
extracted CMDs for stars measured in the field of HS\,85 distributed within the cluster radius (upper-left panel), 
the cluster surrounding field for an equal cluster area (bottom-right panel), and the cluster cleaned from field 
contamination (bottom-left panel). The cluster radial profile is also depicted (upper-right panel).}
\end{figure}

\begin{figure}[htb]
\includegraphics[angle=0,width=17cm]{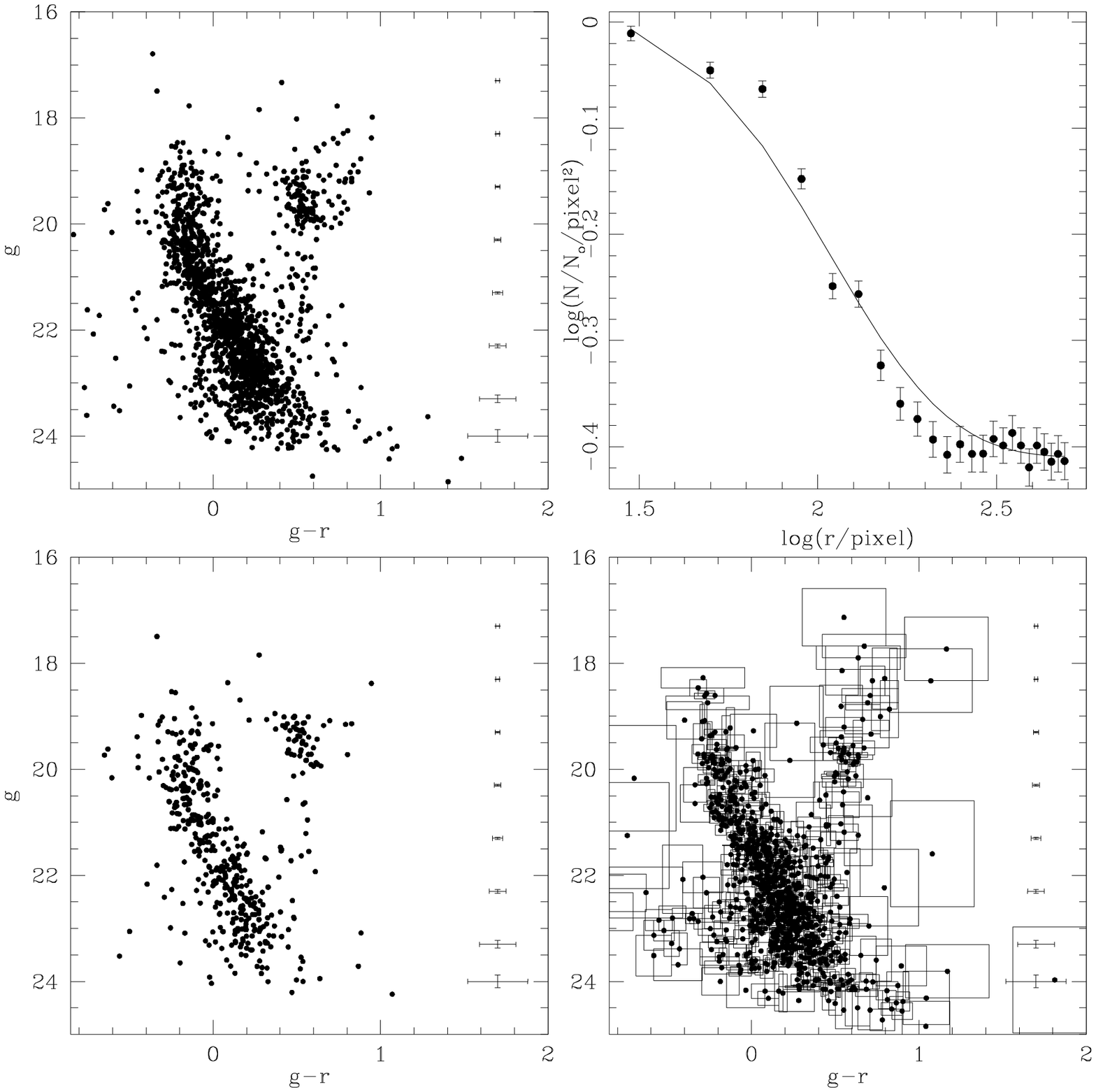}
\caption{Three
extracted CMDs for stars measured in the field of SL\,676 distributed within the cluster radius (upper-left panel), 
the cluster surrounding field for an equal cluster area (bottom-right panel), and the cluster cleaned from field 
contamination (bottom-left panel). The cluster radial profile is also depicted (upper-right panel).}
\end{figure}

\begin{figure}[htb]
\includegraphics[angle=0,width=17cm]{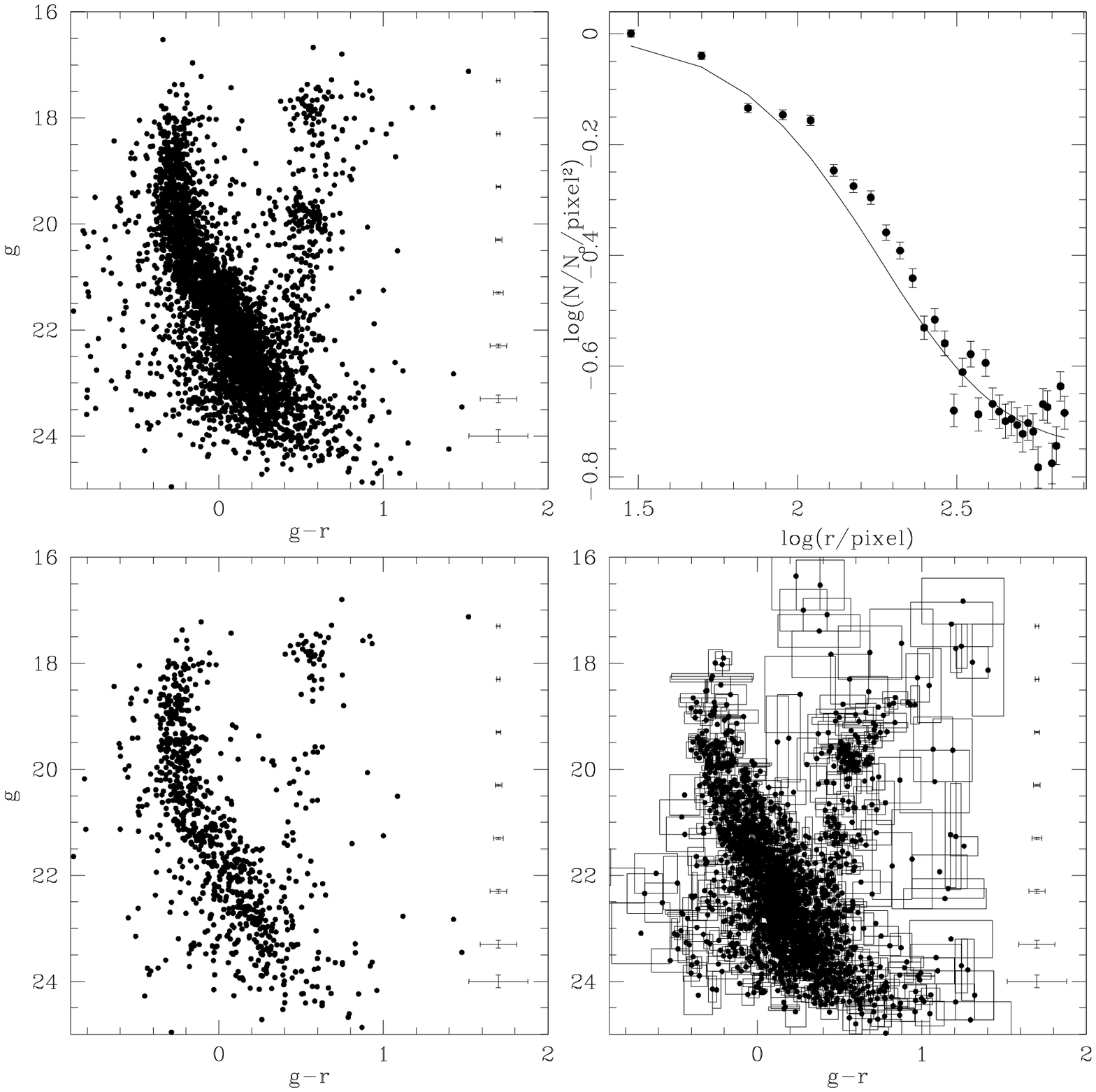}
\caption{Three
extracted CMDs for stars measured in the field of NGC\,2107 distributed within the cluster radius (upper-left panel), 
the cluster surrounding field for an equal cluster area (bottom-right panel), and the cluster cleaned from field 
contamination (bottom-left panel). The cluster radial profile is also depicted (upper-right panel).}
\end{figure}

\begin{figure}[htb]
\includegraphics[angle=0,width=17cm]{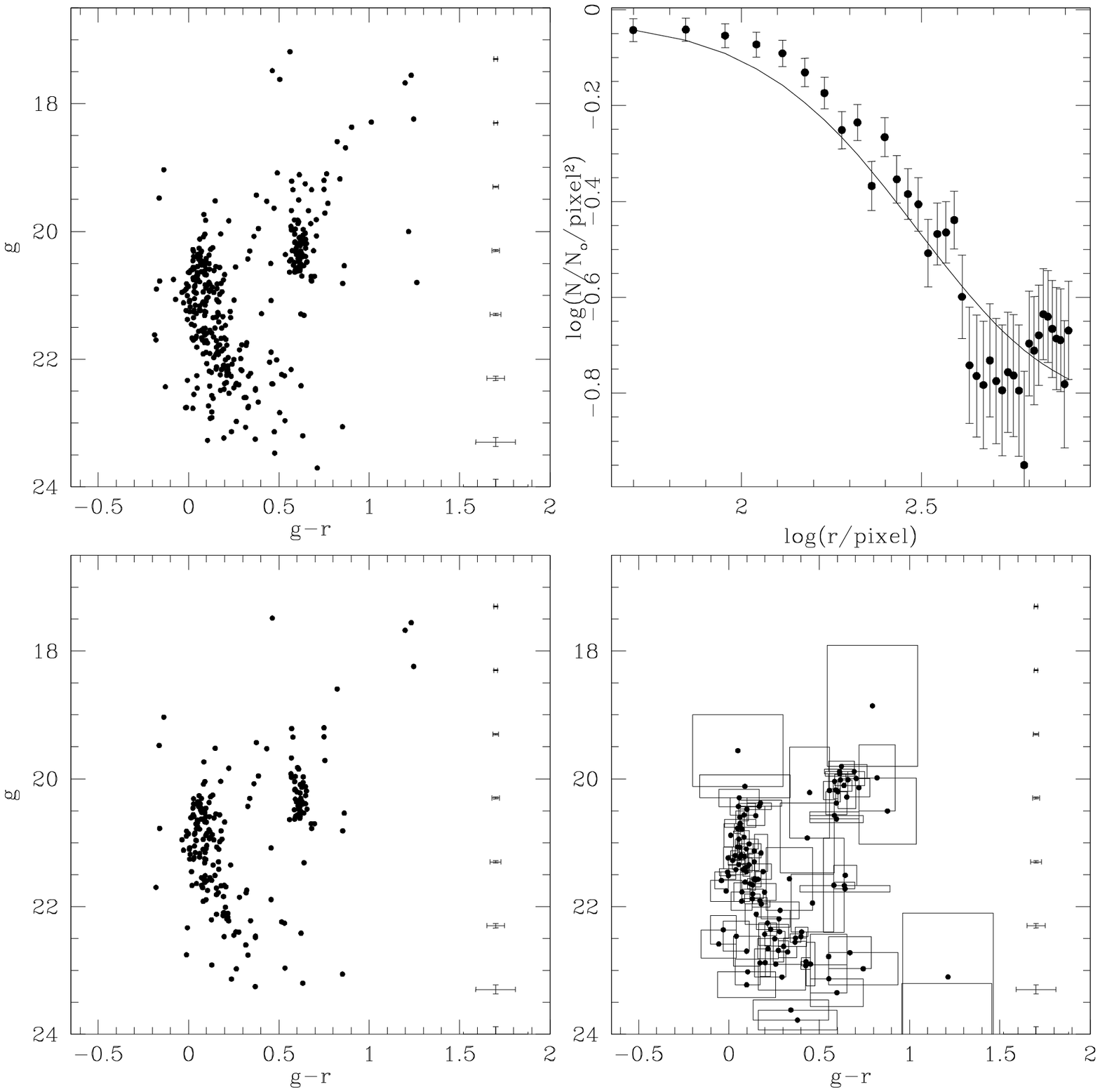}
\caption{Three
extracted CMDs for stars measured in the field of NGC\,2190 distributed within the cluster radius (upper-left panel), 
the cluster surrounding field for an equal cluster area (bottom-right panel), and the cluster cleaned from field 
contamination (bottom-left panel). The cluster radial profile is also depicted (upper-right panel).}
\end{figure}

\begin{figure}[htb]
\includegraphics[angle=0,width=17cm]{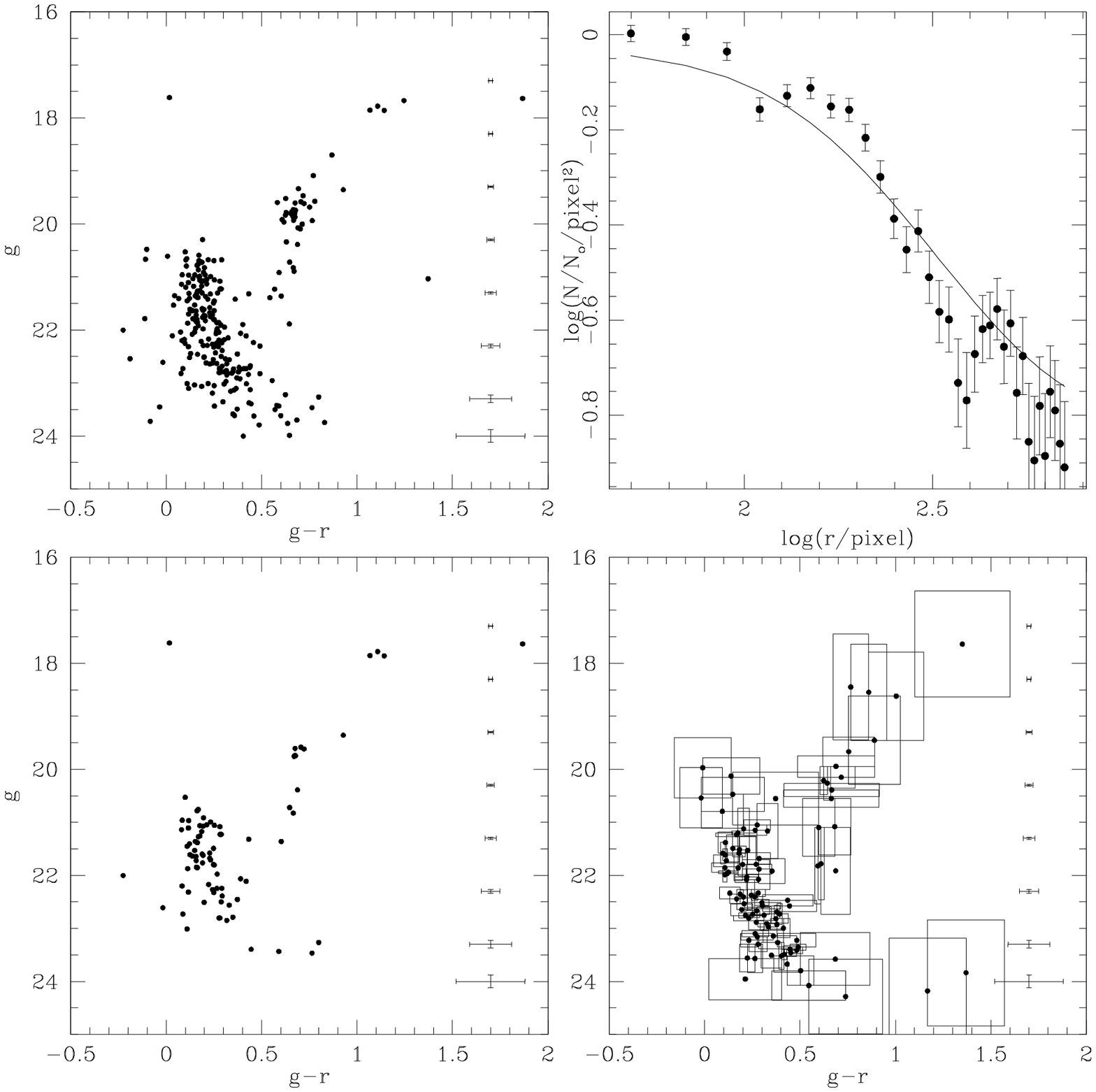}
\caption{Three
extracted CMDs for stars measured in the field of SL\,866 distributed within the cluster radius (upper-left panel), 
the cluster surrounding field for an equal cluster area (bottom-right panel), and the cluster cleaned from field 
contamination (bottom-left panel). The cluster radial profile is also depicted (upper-right panel).}
\end{figure}

\begin{figure}[htb]
\includegraphics[angle=0,width=17cm]{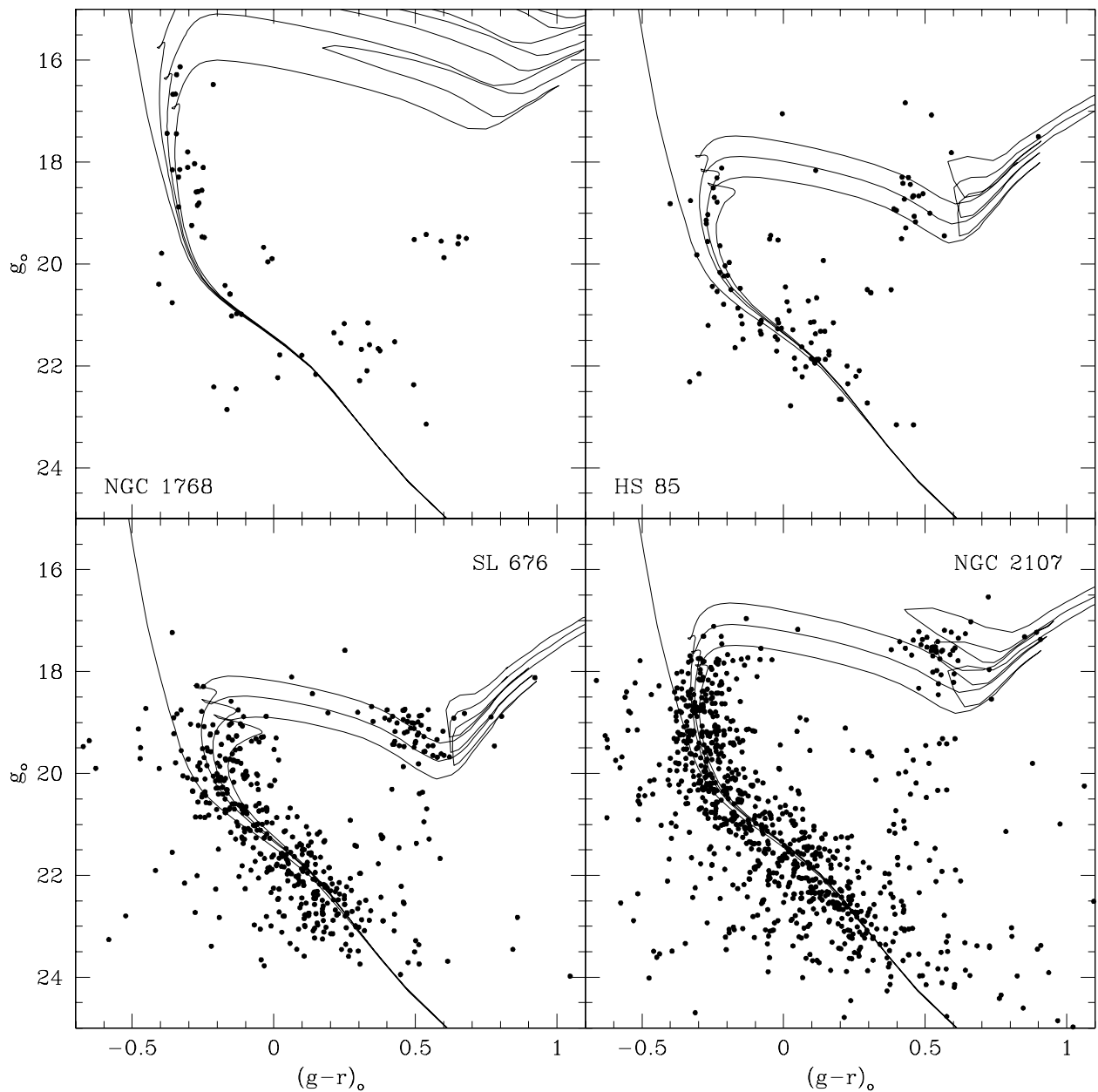}
\caption{The ZAMS and three isochrones (Z = 0.008) from Marigo et al. (2008) superimposed to LMC
cluster CMDs. The youngest isochrone corresponds to log($t$) - $\sigma$(log($t$)) (see Table 8), 
whereas the isochrone separation is $\Delta$(log($t$)) = 0.10.}
\end{figure}

\setcounter{figure}{6}
\begin{figure}[htb]
\includegraphics[angle=0,width=17cm]{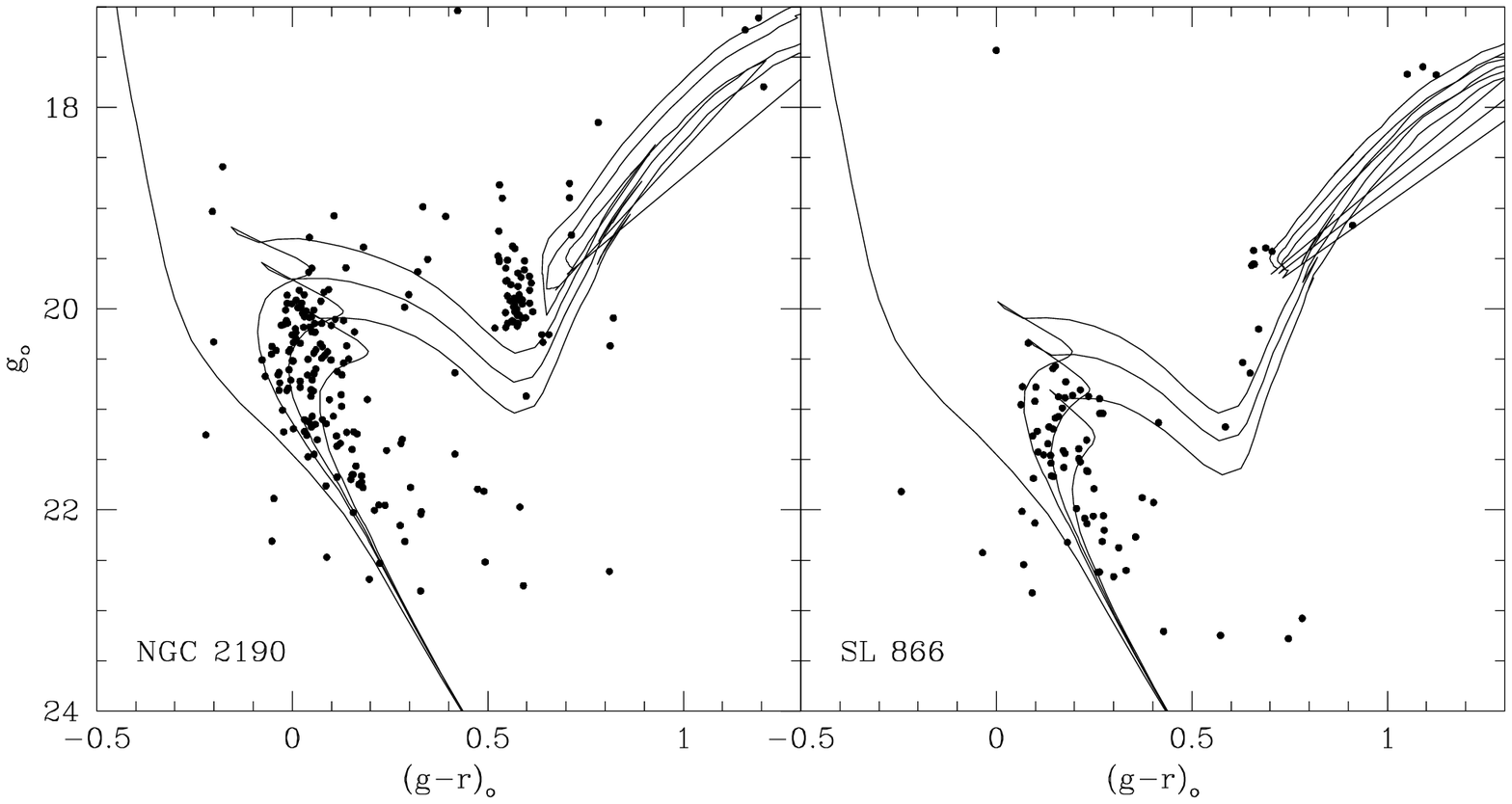}
\caption{continued.}
\end{figure}

\end{document}